\documentclass[aps,prl,twocolumn,epsf,epsfig,amsmath]{revtex4}
\usepackage{latexsym}
\usepackage{hyperref}
\usepackage{times}
\usepackage{graphicx}
\usepackage{amsmath}
\usepackage{multirow} 
\usepackage{dcolumn}
\usepackage{bm}
\usepackage{ textcomp }
\usepackage{color}
\usepackage{amssymb}
\usepackage{xcolor}
\usepackage{easyReview}
\usepackage{mathdots}
\usepackage{float}
\usepackage{lineno}
\usepackage{todonotes}
\usepackage{array}
\usepackage{overpic}

\newcommand{\beq}{\begin{eqnarray}}
\newcommand{\eeq}{\end{eqnarray}}

\newcommand{\refdisp}[1]{Ref. [\onlinecite{#1}]}
\newcommand{\figdisp}[1]{Fig. \ref{#1}}

\begin{document}
\title{Triplet Exciton-driven Topological Mott insulator at Finite Temperature}
\author{Peizhi Mai}
\email{peizhimai@gmail.com}          

\affiliation{Department of Physics and the Anthony J. Leggett Institute of Condensed Matter Theory, University of Illinois at Urbana-Champaign, Urbana, IL 61801, USA}

\begin{abstract}
Motivated by experiments in which the quantum anomalous Hall (QAH) charge gap greatly exceeds the Curie temperature, we apply determinantal quantum Monte Carlo to two complementary lattice models with different geometries: the checkerboard quantum-spin-Hall-Hubbard model and the generalized Kane-Mele-Hubbard model. In both cases an incompressible QAH phase with total Chern number $C=\pm 1$ emerges at quarter filling well above the Curie temperature. Each spin channel carries its own nonquantized Chern number while remaining only partly filled, revealing a topological Mott insulator where Mottness opens the charge gap before magnetic order appears. Charge excitations bind triplet excitons that suppress net magnetization, a many-body dynamical effect absent in mean-field theory. The concurrence of these results on two very different models shows that coupling Mottness with band topology generically yields a high-temperature QAH insulator whose charge excitations are dressed by triplet excitons.
\end{abstract}
\date{today}

\maketitle

\section{Introduction}

In 1988, Haldane pointed out that although quantum Hall topology  requires the breaking of time-reversal symmetry, an external magnetic field is not necessary\cite{HaldanePRL1988}. His lattice model became the prototype for what we now call the quantum anomalous Hall (QAH) effect and sparked an intensive search for materials that would realize it. Experiments lagged behind theory. Only after band theory for time-reversal-invariant topological insulators was established\cite{kanemele1,kanemele2,bhz,MoorePRB2009,FuPRL2007,RoyPRB2009,HasanKaneRMP,QiRMP} and these phases were confirmed experimentally\cite{KonigScience2007,XiaNP2009,ChenScience2009,HsiehNature2008}, did a clear route to QAH emerge: begin with a topological insulator and break time-reversal symmetry. In the absence of an external field, the straightforward path is to induce ferromagnetism, and indeed the first QAH state was achieved in 2013 via magnetic doping\cite{ChangScience2013}.

Recently, the QAH effect has also been observed as an emergent phase produced by the interplay of strong electron correlations and band topology in quantum spin Hall (QSH) systems that respect time-reversal symmetry, with ferromagnetism arising spontaneously from those  correlations. A prominent example occurs at filling factor $\nu=1$ in several moir\'e materials\cite{TingxinLi,FouttyScience2023,YihangZeng,ParkNature2023}. This correlated QAH state is widely regarded as the parent of the fractional QAH phases discovered soon afterwards\cite{YihangZeng,JiaqiCai,XuPRX2023,ParkNature2023}, since both exhibit ferromagnetic order. Our recent numerical simulations demonstrate that this QAH insulator emerges generically from coupling band topology with Mottness, thereby earning the designation “topological Mott insulator” (TMI) \cite{quarter,Haldanequarter}. 


From a phenomenological standpoint, one can also construct such a QAH state  within Hartree-Fock mean-field (MF) theory\cite{SuYing,DongZhihuan,ChangPRB2022,XieYingMing,PanHaining,YangZhang,Devakul,FengchengWu,Rademaker,QiuPRX2023}. MF approaches, however, inherently enforce symmetry breaking and require full spin polarization to obtain a QAH gap. Experiment tells a different story: in moir\'e systems the charge gap opens at relatively high temperature\cite{YihangZeng,RedekopNature2024}, whereas spontaneous symmetry breaking—and certainly full polarization—sets in only at much lower temperatures\cite{YihangZeng,JiaqiCai,XuPRX2023,ParkNature2023}. In other words, the system “knows” it is heading toward a QAH insulator well before ferromagnetism fully develops. This mirrors the classic Mott insulator, where a high-temperature Mott gap precedes low-temperature antiferromagnetic order\cite{ImadaRMP1998}. 

Here we investigate the correlated nature of the QAH state, or TMI, using large-scale, unbiased determinantal quantum Monte Carlo (DQMC) simulations. In the checkerboard-QSH-Hubbard model, we discover that at temperatures well above the onset of ferromagnetism, an incompressible QAH phase emerges at quarter-filling solely from strong electronic correlations.  Both spin species make non-trivial contributions to the Chern number. We further observe that any particle or hole added to this insulator forms a bound state with triplet excitons, underscoring the many body nature of both the gap and the excitation spectrum. To connect with experiments and test generality, we repeat the analysis on the generalized Kane–Mele–Hubbard model on the honeycomb lattice and obtain the same qualitative results. The concurrence across two distinct geometries demonstrates that coupling band topology to Mottness generically yields a correlation-driven QAH insulator without requiring prior ferromagnetism, although magnetic order can still appear at lower temperature. The robust binding of triplet excitons to particle and hole excitations stands out as a clear hallmark of this mechanism.



\section{Results}

We first consider the checkerboard-QSH model\cite{SunPRL2011} with Hubbard interactions in the presence of an external magnetic field. We set the nearest-neighbor hopping scale to unity, $t=1$, and write the Hamiltonian as 

\beq
\begin{aligned}
 H=&-t\sum_{\langle{\bf i}{\bf j}\rangle\sigma} e^{i \phi_{{\bf i},{\bf j}}} e^{\pm i \sigma \psi} 
    c^\dagger_{{\bf i}\sigma}c^{\phantom\dagger}_{{\bf j}\sigma}  - t'\sum_{\langle\langle{\bf i}{\bf j}\rangle\rangle\sigma} (-1)^{\delta_{{\bf i},{\bf j}}} e^{i \phi_{{\bf i},{\bf j}} } c^\dagger_{{\bf i}\sigma}c^{\phantom\dagger}_{{\bf j}\sigma}
    \\&-\mu\sum_{{\bf i},\sigma} n_{{\bf i}\sigma}+ U\sum_{{\bf i}}(n_{{\bf i}\uparrow}-\frac{1}{2})(n_{{\bf i}\downarrow}-\frac{1}{2}),
\end{aligned}
\eeq
where a spin-dependent hopping phase $e^{\pm i \sigma \psi}$ is introduced to ensure opposite Chern number for each spin at half-filling in the absence of the interaction. The sign of hopping phase and the definition of $\delta_{{\bf i},{\bf j}}$ can be found in the supplement. We choose the second-neighbor hopping $t'/t=\sqrt{2}/2$ and $\psi=\pi/4$, following \refdisp{NeupertPRL2011} to have relatively flat topological bands (bandwidth $w/t\approx0.83$ and gap $\Delta/t=4$) and preserve particle-hole symmetry at $\mu=0$, distinct from the generalized Kane-Mele model, which features one flat and one dispersive band. We introduce an external magnetic field through the Peierls substitution as the phase factor $\exp(i \phi_{{\bf i},{\bf j}})$. Here $\phi_{{\bf i},{\bf j}}=(2\pi /\Phi_0) \int_{r_{\bf i}}^{r_{\bf j}} {\bf A}\cdot d{\bf l}$ with the vector potential ${\bf A}=(x\hat{y}-y\hat{x})B/2$. To maintain the single valuedness of the wavefunction, we use the flux quantization condition $\Phi/\Phi_0=n_f/N_c$ which requires a threading per unit cell of  $\Phi=Ba^2$ with $a$ the lattice constant, $n_f$  an integer, and $N_c$ the number of unit cells. The magnetic field serves two purposes: it enables the extraction of the topological invariant (Chern number $C$) via the St$\check{\text{r}}$eda formula\cite{WIDOM1982474,Streda_1982,Streda_1983} and helps reduce the finite-size effects\cite{mfp,quarter,AFM}. We solve this problem using the unbiased DQMC simulations \cite{mfp,Haldanequarter,quarter,AFM,gapopen,DingCP2022,DingPRX2024} on a $6\times6\times2$ cluster. The Jackknife estimate is used to calculate the error bars. Details of the DQMC simulations can be found in the supplement.

\begin{figure}
    \centering
    \begin{overpic}[width=0.5\textwidth, keepaspectratio]{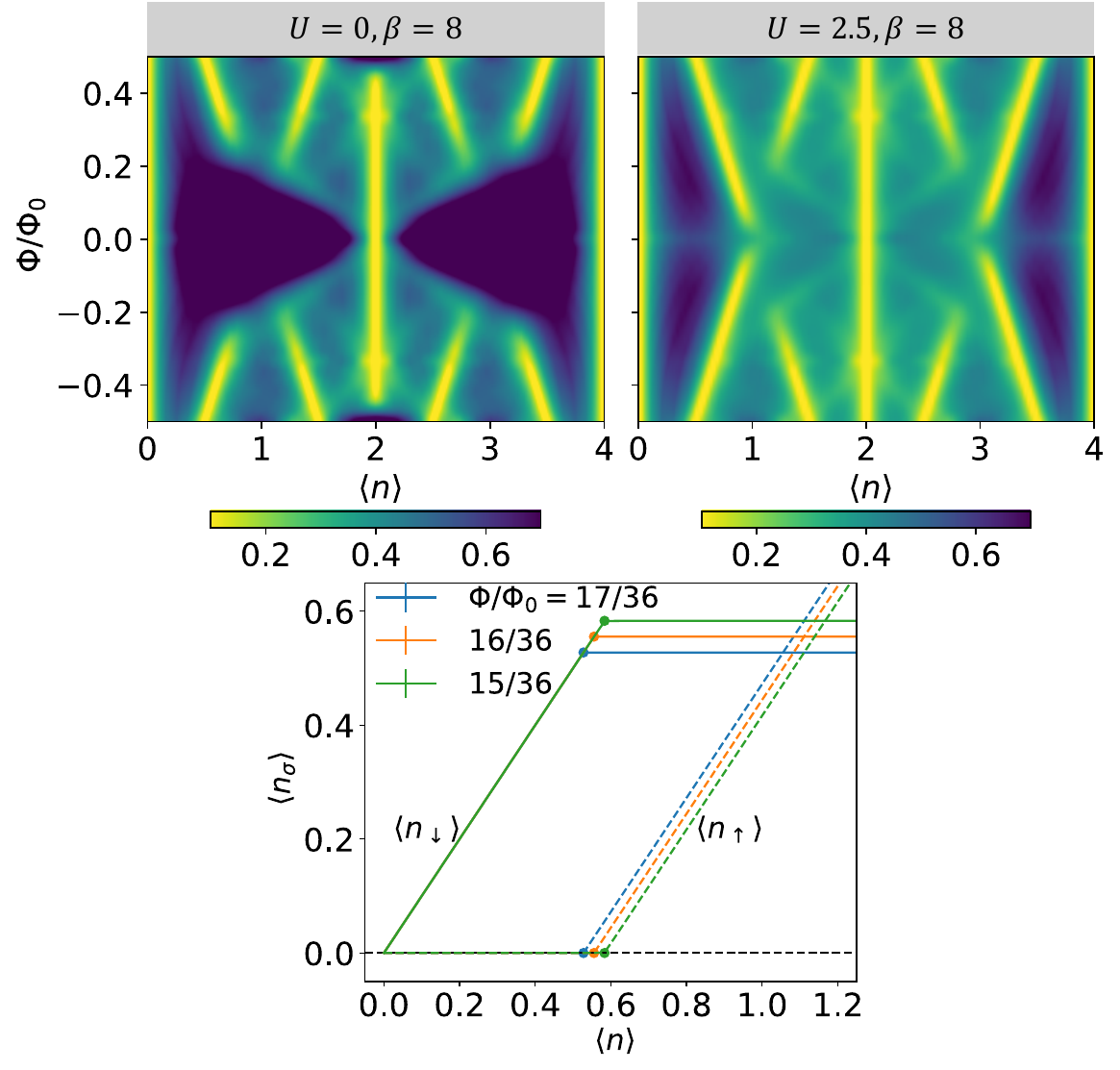}
    \put(21,  92.3){(a)}
    \put(63.5,  92.3){(b)}
    \put(21,  40){(c)}
  \end{overpic}
    \caption{Compressibility $\chi$ as a function of density $\langle n\rangle$ and magnetic flux $\Phi/\Phi_0$ for (a) $U=0$ and (b) $U/t=2.5$. Panel (c) shows the spin-resolved density as a function of total density for high magnetic fluxes at $U=0$. The same high-field results apply to the $U/t=2.5$ case. All panels share the inverse temperature $\beta=8/t$. 
    }
    \label{fig:U0U2p5highf}
\end{figure}

We compute the compressibility 
\beq
\chi=\beta\chi_c=\frac{\beta}{N}\sum_{{\bf i},{\bf j}}\left[ \langle n_{\bf i} n_{\bf j}\rangle - \langle n_{\bf i}\rangle \langle n_{\bf j}\rangle \right], \label{charge}
\eeq
as a function of density $\langle n\rangle$ and magnetic flux $\Phi/\Phi_0$ at $\beta=8t^{-1}$ for $U/t=0$ and $U/t=2.5$ shown in \figdisp{fig:U0U2p5highf} (a) and (b) respectively. The lighter color indicates a dip in the compressibility, signaling an insulator. The inverse slope of the valley corresponds to the Chern number of the incompressible state. In the non-interacting case (\figdisp{fig:U0U2p5highf} (a)), a vertical valley at $\langle n\rangle=2$ gives $C=0$ indicating a quantum spin Hall state at zero field. We already observe a few valleys with finite slope (indicating Chern number $C=\pm 1$) at high field. These Chern insulators are driven by orbital magnetism. Due to the particle-hole symmetry of the system, we can discuss only $\langle n\rangle\leq2$ and the conclusion for $\langle n\rangle\geq2$ is similar after particle-hole transformation. The Chern states at lower density, $\langle n\rangle<1$, appear with full spin polarization. Due to the time-reversal symmetry of the zero-field system, opposite spin has opposite Berry curvature and hence their coupling to the orbital magnetic field leads to splitting of the spin bands even without a Zeeman field, resulting in orbital magnetism. Once the bands for opposite spin fully split at sufficiently high field, filling the lower band leads to an orbital magnetic Chern insulator as shown in \figdisp{fig:U0U2p5highf} (a) with $\langle n\rangle<1$ and $|\Phi/\Phi_0|>0.25$. The $\langle n_\sigma\rangle$ vs $\langle n\rangle$ is also plotted in \figdisp{fig:U0U2p5highf} (c) at high field and low density with markers labeling the corresponding valley in \figdisp{fig:U0U2p5highf} (a). In \figdisp{fig:U0U2p5highf} (c), the system only contains spin-down electrons which contribute to the topology with $C=C_\downarrow=-1$. At higher density $1<\langle n\rangle<1.5$ and high field $\Phi/\Phi_0>0.25$ in \figdisp{fig:U0U2p5highf} (a), the Chern insulator contains both spins which yield $C_\downarrow=-1, C_\uparrow=2$ (details in supplement) and  a total Chern number $C=C_\downarrow+C_\uparrow=1$. In the following, we focus on the lower density case $\langle n\rangle\leq1$ as it is relevant to the response of a topological Mott insulator at $\langle n\rangle=1$ and zero field driven by interaction. We next turn on interaction $U/t=2.5$, which is strongly correlated given the bandwidth $w\approx0.83$ and obtain the compressibility shown in \figdisp{fig:U0U2p5highf} (b). The low-density incompressible Chern insulator initially only occurring at high field extends all the way to zero field. This marks a genuine QAH effect, which we called a TMI. The result matches earlier findings in the interacting Haldane \cite{Haldanequarter} and Kane–Mele \cite{quarter} models, confirming that the combination of Mottness and band topology generically produces such a phase. Although the spin susceptibility shows a peak at $\langle n\rangle=1$ and zero field (see supplement) as in the Kane-Mele-Hubbard case\cite{quarter}, the temperature for \figdisp{fig:U0U2p5highf} (b) is well before any significant spin polarization develops at zero field, yet a clear QAH response is already present. This finding is striking because the prevailing Hartree-Fock MF picture\cite{SuYing,DongZhihuan,ChangPRB2022,XieYingMing,PanHaining,YangZhang,Devakul,FengchengWu,Rademaker,QiuPRX2023} predicts that a QAH gap can form only after the spins become fully polarized. In the following, we present unbiased simulations showing that the gap instead originates from the TMI mechanism, without requiring complete spin alignment, although full polarization does emerge spontaneously at sufficiently low temperature \cite{quarter}.


\begin{figure}
    \centering
    \begin{overpic}[width=0.5\textwidth, keepaspectratio]{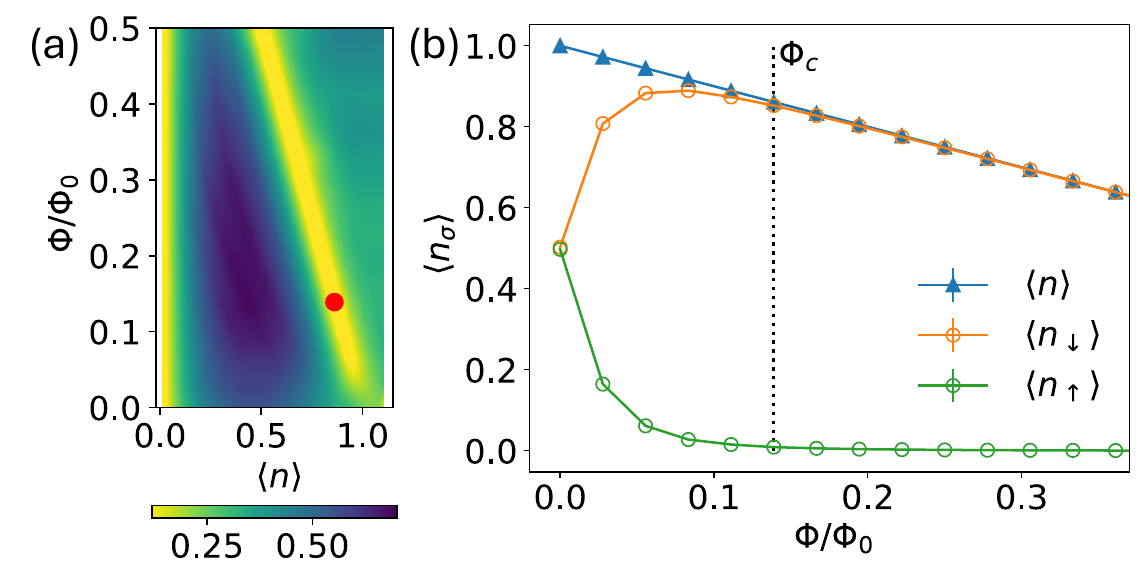}
  \end{overpic}
    \caption{ (a) Compressibility $\chi$ as a function of density  $\langle n\rangle$ and magnetic flux $\Phi/\Phi_0$ at $U=3, \beta=8/t$. (b) Spin-resolved density $\langle n_\sigma\rangle$ and total density $\langle n\rangle$ as a function of magnetic flux along the QAH valley in panel (a). The dotted line labels the critical flux for full spin polarization, providing the red marker in panel (a). 
    }
    \label{fig:U3lowfield}
\end{figure}

In \figdisp{fig:U3lowfield}(a), we plot the compressibility $\chi$ vs $\langle n\rangle$ and positive $\Phi$ at $U/t=3,\beta=8t^{-1}$. As $U=3$ is larger than the value used in \figdisp{fig:U0U2p5highf}(b), the QAH valley is correspondingly deeper, though the temperature remains well above the Curie temperature $\beta_c\approx15$ (see supplement). We show only $\Phi>0$ because the pattern is symmetrical with respect to $\Phi$ (\figdisp{fig:U0U2p5highf}). From the valley region, we extract the total density $\langle n\rangle$ and spin-resolved density $\langle n_\sigma\rangle$, which are conserved because $\hat{S}_z$ symmetry is preserved in the model. These quantities, plotted versus $\Phi/\Phi_0$ in \figdisp{fig:U3lowfield}(b), reveal distinct behaviors. The total density follows $\langle n\rangle=1+C(\Phi/\Phi_0)$ with Chern number $C=-1$. In contrast, the spin-resolved densities $\langle n_\sigma\rangle$ display a richer flux dependence. Above a critical flux $\Phi_c$, defined by the threshold $\langle n_\uparrow\rangle<0.01$, the insulator becomes fully spin polarized and approaches the non-interacting limit observed at high field. For $\Phi<\Phi_c$, an incompressible QAH state still appears despite the presence of both spin species. The nonlinear dependence of $\langle n_\sigma\rangle$ on $\Phi$ implies that the corresponding spin-resolved Chern numbers $C_\sigma=\partial\langle n_\sigma\rangle/\partial(\Phi/\Phi_0)$ are not quantized and also depends on $\Phi$. This means each spin band is only partially filled, while a charge gap has opened, a hallmark of Mott insulator. Moreover, although $\langle n_\uparrow\rangle$ decreases steadily with increasing magnetic field, the majority spin $\langle n_\downarrow\rangle$ varies non-monotonically with $\Phi$, so $C_\downarrow$ flips sign from positive to negative as $\Phi$ grows. In addition, the sharp flux response of $n_\sigma$ at small $\Phi$ shows that $|C_\sigma|$ greatly exceeds $|C|$ in the zero field limit,  yet the total Chern number still satisfies $C=C_\uparrow+C_\downarrow$. In short, even at these relatively high temperatures, the QAH phase forms with both spin channels participating in the topology, each contributing a distinct, non-quantized $C_\sigma$, underscoring the origin being a topological Mott insulator. This mechanism explains why the QAH gap in twisted MoTe$_2$\cite{YihangZeng,RedekopNature2024} far exceeds the material's Curie temperature.


\begin{figure}
    \centering
    \begin{overpic}[width=0.5\textwidth, keepaspectratio]{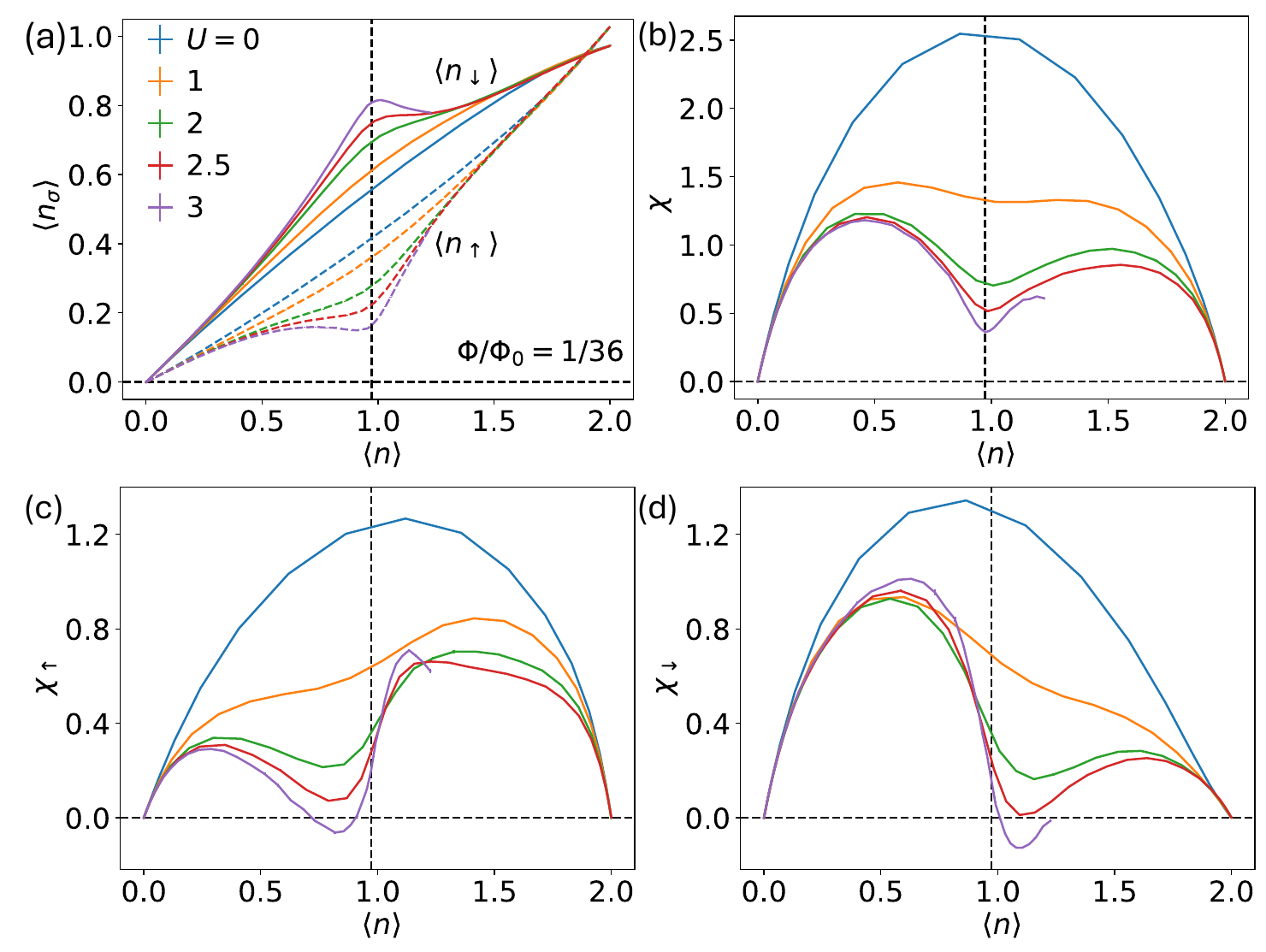}
  \end{overpic}
    \caption{(a-d) Results at the minimal magnetic flux $\Phi/\Phi_0=1/36$ and $\beta=8/t$. (a) Spin-resolved density $\langle n_\sigma\rangle$ versus total density $\langle n\rangle$ with different $U$. Solids lines are for spin-down, while dashed lines represents spin-up electrons. (b) Compressibility $\chi$ and (c,d) Spin-resolved Compressibility $\chi_\sigma$, all versus $\langle n\rangle$ at different $U$. The vertical dashed line signals the insulating QAH density.
    }
    \label{fig:varyU}
\end{figure}

Next we focus on the smallest non-zero magnetic flux $\Phi/\Phi_0=1/36$. Choosing this minimal flux helps in two ways. First, it smooths out finite-size effects\cite{mfp,quarter,Haldanequarter,gapopen,PMQSH}, implying that our results apply to the thermodynamic limit. Second, the weak orbital field slightly offsets the two spin bands; at exactly zero field, DQMC treats the spin channels identically and cannot separate their individual topological responses. The tiny flux therefore allows us to distinguish the spin-resolved contributions and probe the near–zero-field character of the state. We first plot $\langle n_\sigma\rangle$ versus $n$ in \figdisp{fig:varyU}(a) at $\beta=8t^{-1}$ and various values of $U$ (solid lines for $n_\downarrow$ and dashed for $n_\uparrow$). The vertical dashed line marks the QAH density, $\langle n\rangle_{\rm QAH}=1+C(\Phi/\Phi_0)=35/36$. Turning on $U$ amplifies the overall spin imbalance, as expected for a repulsion acting only between opposite spins. The imbalance grows most sharply near $\langle n\rangle_{\rm QAH}$ as $U$ increases. By $U=3$, $\langle n_\downarrow\rangle$ and $\langle n_\uparrow\rangle$ develops a pronounced downturn and upturn, respectively, near the QAH density. This non-monotonic behavior signifies that adding electrons to the QAH state removes carriers from the majority spin and adds carriers to the minority spin; numerically $\delta\langle n_\uparrow\rangle/\delta\langle n\rangle\approx 1.2$ for $U=3t,\beta=8t^{-1}$. In effect, particle or hole excitations bind triplet excitons—a particle-hole pair that carries spin—and the exciton always forms in the spin channel that reduces the net magnetization. 
Fixing $\langle n\rangle=\langle n\rangle_{\rm QAH}$, increasing $U$ enhances ferromagnetism mainly by creating triplet excitons rather than rigidly separating the spin bands; this many-body dynamical mechanism ultimately drives formation of the QAH state. This is one of the key points of this paper. The total charge compressibility in \figdisp{fig:varyU}(b) develops a pronounced dip as $U$ increases, signaling the incompressible QAH phase driven by Mottness. Motivated by the spin-resolved density analysis, we also examine the spin-resolved compressibility $\chi_\sigma=\partial\langle n_\sigma\rangle/\partial\mu$, shown in \figdisp{fig:varyU}(c,d). As $U$ increases, $\chi_\uparrow$ becomes negative just below the QAH filling, while $\chi_\downarrow$ becomes negative just above it. These negative regions originate from the non-monotonic features in \figdisp{fig:varyU}(a) and are direct signatures of exciton binding to particle or hole excitations. These are dynamical correlation effects that cannot be captured by a single-particle MF framework. Although Mottness greatly enhances magnetization exactly at the QAH density, the system quickly reverts to ordinary orbital behavior once the density is shifted due to the population of excitons carrying the opposite spin.

\begin{figure}
    \centering
    \begin{overpic}[width=0.5\textwidth, keepaspectratio]{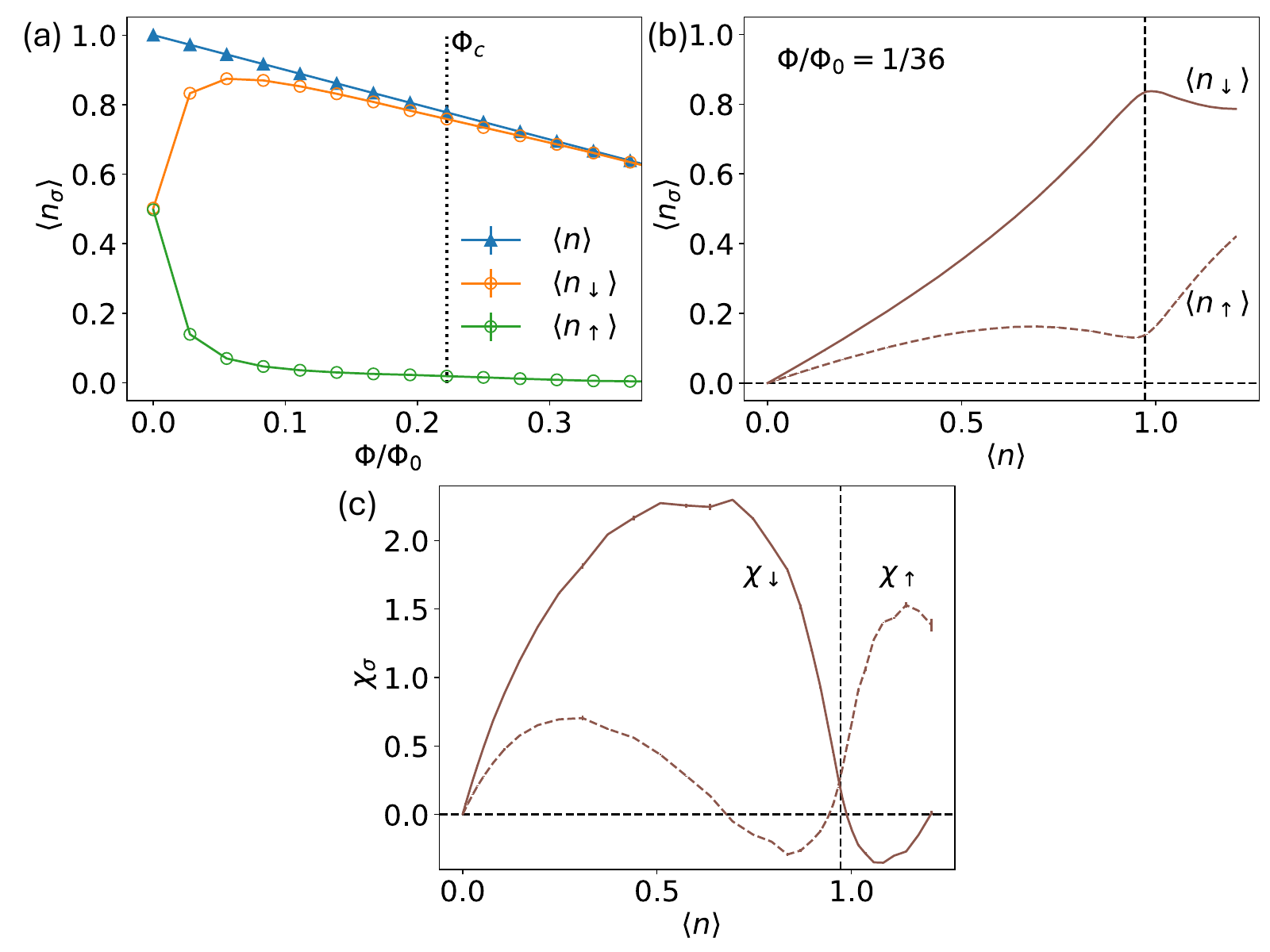}
  \end{overpic}
    \caption{(a) Spin-resolved density $\langle n_\sigma\rangle$ and total density $\langle n\rangle$ versus magnetic flux $\Phi/\Phi_0$ along the QAH valley. The dotted line labels the critical flux for full spin polarization. The Panels (b) and (c) are Spin-resolved density $\langle n_\sigma\rangle$ and compressibility $\chi_\sigma$, respectively, both as a function of total density $\langle n\rangle$ at minimal magnetic flux. The dashed line signals the insulating QAH density. All panels are simulations from the Kane-Mele-Hubbard model at $U/t=1.5,\beta=16/t$.
    }
    \label{fig:KMH}
\end{figure}

The checkerboard-QSH-Hubbard model offers a clean starting point with both topological bands flat and particle-hole symmetry at half-filling. To make closer contact with the moir\'e TMD materials\cite{Devakul,FengchengWu,DevakulPRX2022}, we now turn to the generalized Kane-Mele-Hubbard model in a magnetic field. The Hamiltonian is \cite{quarter}

\beq
\begin{aligned}
\label{Eq:KMBfield}
    H_{\text{KM}}=&-t\sum_{\langle{\bf i}{\bf j}\rangle\sigma} e^{i \phi_{{\bf i},{\bf j}}} 
    c^\dagger_{{\bf i}\sigma}c^{\phantom\dagger}_{{\bf j}\sigma}  -t'\sum_{\langle\langle{\bf i}{\bf j}\rangle\rangle\sigma}e^{\pm i \psi\sigma} e^{i \phi_{{\bf i},{\bf j}} } 
    c^\dagger_{{\bf i}\sigma}c^{\phantom\dagger}_{{\bf j}\sigma}\\&-\mu\sum_{{\bf i},\sigma} n_{{\bf i}\sigma}+ U\sum_{{\bf i}}(n_{{\bf i}\uparrow}-\frac{1}{2})(n_{{\bf i}\downarrow}-\frac{1}{2}),
\end{aligned}
\eeq
We choose $t'=0.3t,\psi=0.81\pi$. At $U=0$, the lower band is nearly flat with bandwidth $w_\text{lower}\approx0.28t$; the upper band is highly dispersive with $w_\text{upper}\approx4.36t$; and the two are separated by a band gap $\Delta\approx1.63t$. Given the small bandwidth of the lower band, a moderate $U=1.5t$ will suffice to drive it into a strongly correlated regime. Working at the lower temperature $\beta=16t^{-1}$, we present the spin-resolved density $\langle n_\sigma\rangle$ and compressibility $\chi_\sigma$ in \figdisp{fig:KMH}. The results essentially mirror what we found for the checkerboard-QSH-Hubbard model in \figdisp{fig:U3lowfield} and \figdisp{fig:varyU}. In \figdisp{fig:KMH}(a), the QAH state again forms while both spins are present; for $\Phi<\Phi_c$, each spin carries a non-quantized Chern number, with $|C_\sigma|\gg1$ as $\Phi\rightarrow0$, while $C=C_\uparrow+C_\downarrow=-1$ for all $\Phi$. \figdisp{fig:KMH}(b) and (c) shows that the non-monotonic filling dependence of $\langle n_\sigma\rangle$ ($\delta\langle n_\uparrow\rangle/\delta\langle n\rangle\approx 1.3$ for electron addition) and the associated negative spin-resolved compressibility $\chi_\sigma$ persist in this model as well. These features indicate that charge excitations tend to bind with triplet excitons that counteract the magnetization. The same signatures including non-quantized spin-resolved Chern numbers, negative $\chi_\sigma$, and spin-selective exciton binding, appear in both lattice geometries, demonstrating that these features are generic hallmarks of the topological Mott insulator.


\section{Discussion}
We used unbiased determinantal quantum Monte Carlo to study the checkerboard QSH-Hubbard model and the generalized Kane–Mele Hubbard model. In both systems, we find that well above the Curie temperature an incompressible QAH phase appears with total Chern number $C=\pm1$. Each spin channel contributes a non-quantized Chern number while remaining only partially filled, a signature of a topological Mott insulator: Mottness opens the charge gap in partially filled bands first, and magnetic order follows only at lower temperature. This matches what is seen in twisted MoTe$_2$\cite{YihangZeng,RedekopNature2024} and suggests a similar sequence for other interaction-driven topological gaps. In particular, the charge gap of the recently reported high-temperature fractional Chern insulator\cite{Parkarxiv2025} should open first, with magnetic order appearing only at lower temperature. Moreover, we observe that charge excitations in the QAH state are dressed by triplet excitons that reduce the net magnetization. This indicates that a finite population of triplet excitons plays a central role in stabilizing the QAH state. As this population increases at lower temperature or stronger interaction, these dressed charges may evolve into true trions. 
This generic mechanism where a finite-density of triplet exciton driven by interaction stabilizes the QAH effect and dresses its charge excitations differs from recent proposals\cite{DongZhihuan,ZhenNC2021,XiePRB2024} on excitonic Chern insulator which invoke spin-singlet excitons produced by long-range Coulomb repulsion\cite{ZhenNC2021}, an external displacement field\cite{DongZhihuan} or Hund's coupling\cite{XiePRB2024}. Their presence provides an additional hallmark of the topological Mott insulator. Because these effects arise from many-body correlations, they lie beyond any Hartree–Fock description and emerge only when interactions are treated without bias and full dynamics are retained. The fact that two very different lattice geometries yield the same physics underscores that the interplay between topology and strong correlations is generic, not material specific. 

\section{Acknowledgements}

We thank Shizeng Lin for suggesting the comparison between MF ferromagnetic QAH and TMI. We also thank Andrea Young, Philip Phillips, Liang Fu, Yihang Zeng, Jiaqi Cai and Taige Wang for insightful discussions. This work was supported by the Gordon and Betty Moore Foundation’s EPiQS Initiative through grant GBMF 8691. The DQMC simulations of this work used the Advanced Cyberinfrastructure Coordination Ecosystem: Services \& Support (ACCESS) Expanse supercomputer through the research allocation TG-PHY220042, which is supported by National Science Foundation grant number ACI-1548562\cite{xsede}.

\bibliography{reference}

\end{document}